# Reply to Comment on "Isotope Effect in High-$T_C$ Superconductors"

Dale R. Harshman

*Physikon Research Corporation, Lynden, Washington 98264, USA;*
*Department of Physics, University of Notre Dame, Notre Dame, Indiana 46556, USA;*
*and Department of Physics, Arizona State University, Tempe, Arizona 85287, USA*

John D. Dow

*Department of Physics, Arizona State University, Tempe, Arizona 85287, USA;*
*and Institute for Postdoctoral Studies, 6031 East Cholla Lane, Scottsdale, AZ 85253 USA*

Anthony T. Fiory

*Department of Physics, New Jersey Institute of Technology, Newark, New Jersey 07102, USA*

## ABSTRACT

Our paper on the isotope effect in high-temperature superconductors with cation substitutions presents a comprehensive analysis rooted completely in the experimental evidence. In this Reply we show that pair-breaking disorder, isotope effects, doping-induced variations in $T_C$ and in the magnetic penetration depth, Coulomb's law and Anderson's theorem are treated with correct physical and mathematical fundamentals. In contrast, the theory fostered in the Comment by Alexandrov and Zhao contradicts several specific experimental facts, eight of which are briefly discussed. Their Comment also uncritically repeats a previously discredited assertion of an isotope effect in the superconducting carrier mass, incorrectly assumes that cation doping continuously varies *intrinsic* superconducting parameters, unjustifiably assigns importance to data from samples with serious quality problems, and renders a false estimate of the pair-breaking strength.

**Keywords:** High-$T_C$ Superconductivity, Isotope Effect, Optimization, Pair Breaking

**PACS numbers:** 74.20.-z, 74.25.Kc, 74.72.-h, 74.70.Kn, 71.38.-k

## I. Introduction

Our paper on the isotope effect in high-$T_C$ superconductors discusses compounds that are modified by cation doping, which invariably reduces the transition temperature [1]. Our interpretation follows from the general observation that the *doped materials are not intrinsic superconductors* because their key characteristic is inhomogeneous superconductivity. In contrast, the criticisms leveled against our work in the Comment by Alexandrov and Zhao are based on treating such doped compounds as intrinsic superconductors, albeit with lower transition temperatures [2], a notion which has been thoroughly discredited [3]. In Section II of this Reply we explain why Alexandrov's and Zhao's assertions of "violating Anderson's theorem and the Coulomb law" and their criticisms of our mathematics and physics are incorrect in their entirety. Given the abundant discussion in the Comment concerning polaronic and phononic models, it becomes necessary to summarize the overwhelming contrary experimental evidence in Section III. Section IV summarizes the relationships between our responses to the Alexandrov-Zhao criticisms and the errors in their Comment. We introduce the responses in our Reply first by briefly enumerating the relevant experimental facts:

1. In the case of the oxygen isotope effect (OIE) in $T_C$, whose mass dependence is $T_C \sim M^{-\alpha_O}$, the exponent $\alpha_O$ is vanishingly small for $YBa_2Cu_3O_{6.95}$ ($\alpha_O \leq 0.03$) when compared to the BCS (Bardeen-Cooper-Schrieffer) value (i.e., $\alpha_O = 0.5$), but increases as one moves away from optimal stoichiometry and as $T_C$ decreases. Figures 1 and 2 show the strong correlations of $\alpha_O$ with transition width $\Delta T_C$ and Meissner fraction, respectively, for the case of Pr substituted $YBa_2Cu_3O_{7-\delta}$ [4,5], indicating a dependence of the OIE on sample quality.

2. It is well established that samples exhibiting excessively broadened superconducting transitions, $\Delta T_C$, and degraded diamagnetic screening (i.e., a diminished Meissner effect), do not contain a *homogeneous superconducting state* (see Figs. 1 and 2), and hence make the data acquired for such samples suspect.

3. Experiments find no change of the OIE in the penetration depth, $\delta\lambda_{ab}/\lambda_{ab}$ (i.e., the fractional change in penetration depth upon oxygen isotope substitution), when $T_C$ is depressed by cation-substitution disorder and $\alpha_O$ increases by an order of magnitude (see Fig. 3 of Ref. [1]). This behavior is also shown in Fig. 3 herein for the Pr-substituted $YBa_2Cu_3O_{7-\delta}$, where $\delta\lambda_{ab}/\lambda_{ab}$ remains constant, with the only exceptions being for samples with the largest broadening ($\Delta T_C \sim 9$ K) [4,6].

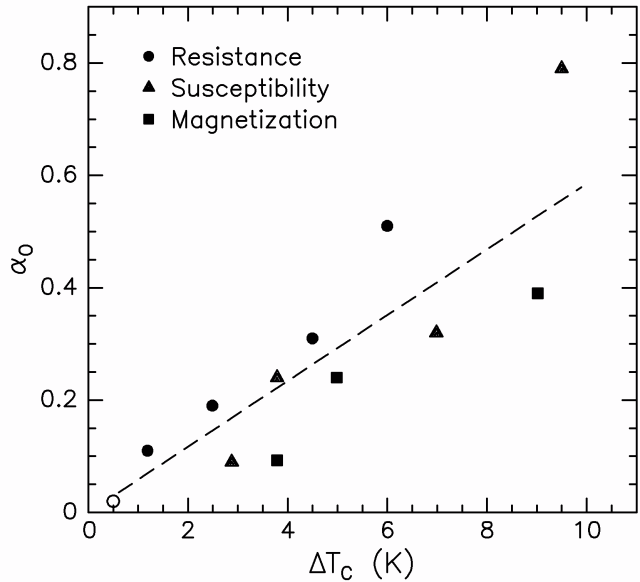

**Figure 1.** Exponent of the oxygen isotope effect in the superconducting transition, $\alpha_O$, plotted against the width of the superconducting transition, $\Delta T_C$, for Pr-doped $YBa_2Cu_3O_{7-\delta}$, as determined from measurements of resistance, susceptibility, and magnetization [4]. The open symbol is for undoped $YBa_2Cu_3O_{7-\delta}$. The dashed line denotes the trend.

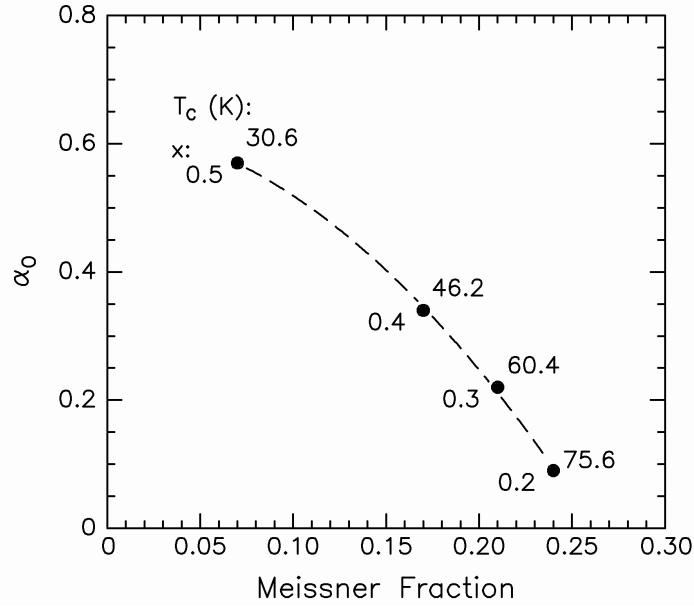

**Figure 2.** Exponent of the oxygen isotope effect in the superconducting transition, $\alpha_O$, plotted against the Meissner-effect fraction for Pr-doped $YBa_2Cu_3O_{7-\delta}$ [5] (fraction is unity when undoped). Labels denote $T_C$ and Pr fraction x per formula unit for each datum. The dashed curve is a guide to the eye.

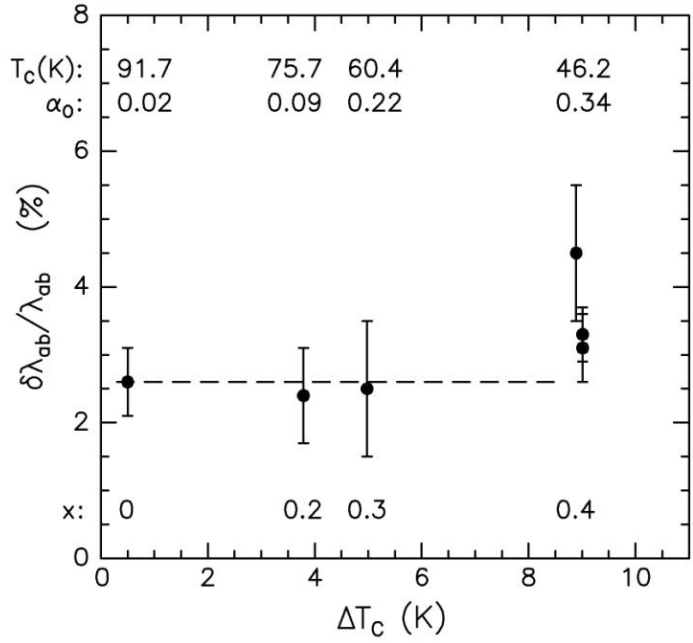

**Figure 3.** Oxygen isotope effect in the superconducting magnetic penetration depth [6], $\delta\lambda_{ab}/\lambda_{ab}$, plotted against the width of the superonducting transition, $\Delta T_C$ (from magnetization) [4] for Pr-doped $YBa_2Cu_3O_{7-\delta}$. Labels denote $T_C$, $\alpha_O$, and the Pr fraction $x$ per formula unit for each datum. For $\Delta T_C < 9$ K ($x < 0.4$) there is no observable change of the OIE in the magnetic penetration depth (dashed line); only for $\Delta T_C > 9$ K ($x > 0.4$) is a change observable. See also Fig. 3 of Ref. [1].

4. Doping with Pr on the Ba site dramatically depresses $T_C$ and broadens the superconducting transition (see Fig. 1), whereas doping with Pr on the Y site has little effect on $T_C$. Experiment thus reveals the crucial role played by the Ba site for superconductivity (see Ref. [1], and references cited therein).

5. A critical observation is the invariance of $T_C$ for up to 96.8% increases in the mass of the atom at the Y site in $YBa_2Cu_3O_{6.95}$ (for substitutions from Y to Lu). This *zero mass dependence* of $T_C$ is unexplained by models based on pairing via lattice vibrations [1].

6. Evidence against lattice-phonon involvement in the pairing mechanism is the absence of case II coherence factor effects, e.g., no Hebel-Slichter anomaly in nuclear magnetic resonance (NMR) [7,8].

7. All phonon-mediated superconductors display a strong non-linear behavior of the normal-state resistivity $\rho(T)$ which saturates at elevated temperatures, while high-$T_C$ materials display an absence of saturation and linearity in $\rho(T)$ from $T_C$ to high temperatures. Consequently, one can rightly deduce, as reviewed in Ref. [1] and confirmed by first principles calculations, a small electron-phonon coupling parameter $\lambda \sim 0.27$ and $T_C \leq 2$ K [9].

8. Bulk probes of the superconducting condensate provide a consistent interpretation of a nodeless pairing state with extremely strong coupling [8][10-15]. For non-bulk probes, as observed by G. M. Zhao (co-author of the Comment) [16], *"...phase sensitive experiments are probing the OP* [order parameter] *symmetry on surfaces and interfaces, which are found to be significantly underdoped,"* and concludes that, *"...these surface- and phase-sensitive experiments do not provide conclusive evidence for d-wave gap symmetry in the bulk of high-temperature superconductors."* For example, Zhao finds that the gap in $Bi_2Sr_2CaCu_2O_{8+\delta}$ is nodeless along the diagonal direction, ruling out *d*-wave symmetry [16], which is also the case for $YBa_2Cu_3O_{7-\delta}$ from c-axis tunneling experiments [17].

## II. Responses to Criticisms

Here we show that correct mathematical models for the pair-breaking effect on the transition temperature and the penetration depth are used, and no violation of Anderson's theorem or of Coulomb's law occurs in our work [1].

### A. Pair breaking: presumed violation of Anderson's theorem

The assertion about "violating Anderson's theorem" is predicated on our treatment of the pair-breaking effect of non-magnetic impurities, which the authors of the Comment [2] mistakenly consider solely in terms of gap symmetry. It is of great importance to recognize that discussions of the parameter *a* in Eq. (1) of Ref. [2] and the various analyses derived therefrom refer to theories of pair breaking that do not include the behavior of inhomogeneous superconductors with strongly two-dimensional electronic character (see Sect. I, facts #1-#4). Alexandrov and Zhao [2] do note the constraints of the Anderson theorem, in which the transition temperature is unaffected by conventional impurity scattering in phonon-mediated superconductors. For example, the transition temperature of an alloy, such as $Pb_{1-x}In_x$, depends only on the electron-phonon coupling and not on electron-impurity scattering induced by alloying. This is what the authors of Ref. [2] assume in *their* model. However, owing to facts #1-#4 regarding alloying-induced inhomogeneity in the high-$T_C$ materials, the cation-substituted high-$T_C$ superconductors ought not be treated as conventional homogeneously alloyed superconductors (e.g., like $Pb_{1-x}In_x$). In general, Anderson's theorem applies to homogeneous superconductors; it by itself does not wholly describe inhomogeneous superconductors.

The formulation we present in Ref. [1] properly considers the experimental facts, our understanding of quasi two-dimensional electronic transport in $YBa_2Cu_3O_{6.95}$, and the accepted theory for the effect of disorder on $T_C$ in two-dimensional superconductors [18]. The theory was experimentally validated for thin films of *s*-wave superconductors in Ref. [18] (which also cites the substantial body of theoretical background). Our formula (same as the Abrikosov-Gorkov expression with $a = 1$ and introduced by Kresin et al. [19][20]) correctly describes the relevant physical mechanism, which is pair breaking in inhomogeneous superconductors with disorder and carrier interaction effects in two dimensions, and is applied in a manner consistent with Anderson's theorem. In particular, the effect of Pr-doping mimics the behavior of thin films, wherein disorder depresses $T_C$ and broadens the superconducting transition. The intimate connection between the exponent $\alpha_O$ and $\Delta T_C$ that is illustrated in Fig. 1 shows that our physical model is the correct one. The resulting theory [1], wherein $\alpha_O$ scales with the pair-breaking parameter, is simple and removes the unphysical singularity at optimal $T_{C0}$ present in previous models [6][19].

Reference [2] incorrectly asserts that the pair-breaking parameter is small, by estimating it from the scattering rate in $YBa_2Cu_3O_{6.5}$ and then assuming the same result applies to cation doping. Since $YBa_2Cu_3O_{6.5}$ is near the oxygen-ordered region with $T_{C0} \sim 60$ K, it is in fact a near-optimum compound, possessing an unrepresentatively low scattering rate. One errs in misrepresenting it as a case of disorder, since removal of $O^{-2}$ *anions* does not directly mirror *cation* doping. Alexandrov and Zhao also fail to address the inhomogeneous or percolative conduction that is characteristic of cation-substituted samples [19], which prevents quantitative analysis. Thus, in cases of inhomogeneous superconductivity, scattering estimates such as that claimed in the Comment do not reveal the presence or strength of a pair-breaking effect on $T_C$.

## B. Isotope effect in the penetration depth

The Comment's proposed alternative formulation for the effect of pair breaking on the penetration depth leads to disagreement with OIE experiments (see fact #3). For example, the fractional change in penetration depth $\delta\lambda_{ab}/\lambda_{ab}$ for Pr-substituted $YBa_2Cu_3O_{7-\delta}$ (Fig. 3) remains statistically constant with doping up to Pr fraction $x = 0.4$ [4,6], even though $\alpha_O$ increases substantially, as shown by labels in Fig. 3. At $x = 0.4$ the transition width has broadened to $\Delta T_C/T_C = 20\%$ (Figs. 2 and 3) and the Meissner fraction has dropped to a mere 17% (Fig. 2). (Note that at $x = 0.5$ (see Fig. 2), i.e., where the inhomogeneous superconductivity itself is on the verge of disappearing, there are no data on the OIE in penetration depth.) Experiments thus disprove two theses in the Comment: enhancement of the OIE in $\lambda_{ab}$ (which is negligibly small) and its similarity to the OIE in $T_C$ (which is comparatively huge) as one varies cation doping. On the other hand, experiments do validate the formulation of Ref. [1], which uses $T_{C0}$ in the right hand side of the expression for the penetration depth,

$$\lambda_{ab}^2(T_C) = \lambda_{ab}^2(T_{C0}) \left[1 + \eta \tilde{\alpha}/k_B T_{C0}\right], \qquad (1)$$

where $\eta = 0.36$. As discussed in Ref. 1, our Eq.(1) predicts the negligible variation ($\pm 0.1\%$ and smaller than error bars in Fig. 3) in $\delta\lambda_{ab}/\lambda_{ab}$, in agreement with the observed invariance.

We understand the correctness of Eq. (1) as follows: Increasing the Pr-on-Ba-site defect concentration in Pr-substituted $YBa_2Cu_3O_{6.95}$ has the effect of degrading the superconducting phase's transition temperature, broadening it, and diminishing the Meissner effect (Figs. 1 and 2). We assert that only the parent compound $YBa_2Cu_3O_{6.95}$ may be considered as an intrinsic superconductor, i.e., with intrinsic attributes of transition temperature $T_{C0}$, London penetration depth $\lambda_L$, and Pippard coherence distance $\xi_0$. These doped materials must therefore be deemed inhomogeneous *extrinsic derivatives* of the intrinsic superconductor, with lower transition temperatures $T_C < T_{C0}$, but have essentially the same underlying intrinsic $T_{C0}$, $\lambda_L$, and $\xi_0$ parameters, at least in the fraction of material (e.g., Meissner fraction, Fig. 2) that remains superconducting. By blindly following the conventions associated with materials like $Pb_{1-x}In_x$ (in which $T_C$ is not suppressed by impurity scattering), and inserting $T_C$ in place of $T_{C0}$ in the second term of Eq. (1) above, Alexandrov and Zhao continue to ignore the sample degradation which accompanies cation doping in the high-$T_C$ compounds. Cation-doped materials cannot be equated with conventional homogeneous superconductors with magnetic impurities, in which case one does use $T_C$ [21]. Our version of Eq. (1) better approximates the physical nature of such doping on the bulk high-$T_C$ superconducting state and is in excellent agreement with the observed invariance for $-\delta T_C/T_C \leq 3\%$ as shown in Fig. 3 of Ref. [1] (our original work). Beyond this doping level the quality of the materials becomes truly problematic.

### C. Presumed violation of Coulomb's law

The "Coulomb's law" critique argues that $\alpha_O$ is a signature of high-frequency vibrational modes that are unscreened in the parent compound. By logical extension of this argument these modes would be unscreened in the doped compounds as well, and, in contradiction to experiment, $\alpha_O$ would remain nearly constant. The predicted $\alpha_O$ also would probably be much larger than it is for $YBa_2Cu_3O_{6.95}$ ($\alpha_O \leq 0.03$, see Fig. 1). Since the authors' critique hinges on unsound theory, there is no violation of Coulomb's law.

### III. Failures of Polaron Theory

Herein we discuss four basic shortcomings of the Fröhlich electron-phonon interaction model of hole pairing advocated by Alexandrov and Zhao [2].

### A. Mass substitutions and the OIE

Our conclusion that phonon interactions are insufficient for the superconductive pairing mechanism is a straightforward recognition that the effects of mass substitutions on $T_{C0}$ are small or non-existent (Sect. I, facts #1 and #5), that case II coherence factors are absent (fact #6), and that estimates and calculations show $\lambda$ to be small (fact #7). For example, the shift in $T_{C0}$ with $O^{16} \to O^{18}$ substitution is $-0.3\%$ or less in magnitude for optimum $YBa_2Cu_3O_{6.95}$ and $-1.4\%$ for near optimum $La_{2-x}Sr_xCuO_4$ [1]. For Y→Lu substitution no change from $T_{C0}$ is observed. Yet shifts in phonon frequencies are expected in each case, irrespective of whether a particular ion is contributing carriers to the superconducting condensate. From a reduced mass model for optical vibrations of $YCu_2O_4$ or $YBa_2Cu_2O_6$ substructures [22], we estimate OIE frequency shifts of $-2.5\%$ and $-1.8\%$, respectively. For Y→Lu substitution (recalling that the Y site is ~1.6 Å from the $CuO_2$ layers) the corresponding shifts in frequency are $-8.3\%$ and $-8.7\%$. Since these frequency shifts greatly exceed the observed shift in $T_{C0}$ for $YBa_2Cu_3O_{6.95}$, it becomes clear that experiment renders the Fröhlich electron-phonon interaction model of hole pairing advocated by Alexandrov and Zhao [2] self-inconsistent. To buttress their view, the authors of the Comment cite model exercises that predict electron-phonon interactions to be large, while other first principles calculations that predict otherwise remain uncited [9].

### B. Absence of OIE in carrier mass

An OIE in the superconducting carrier mass (i.e., dependence of the carrier mass on oxygen isotope), which is an essential feature of polaron theory, is claimed to exist by Alexandrov and Zhao [2]. Unfortunately for the viability of polaron theory, the arguments presented in Ref. [2] supporting variation of carrier mass with oxygen isotope repeat an erroneous interpretation of the OIE in the penetration depth that can be traced back to earlier experiments [23]. Reference [23] argues that the OIE in $\lambda_{ab}$ is instead associated with the carrier density. Whatever the truth of the matter is, neither the authors of the Comment [2] nor any of those they cite have acknowledged the fact that an OIE in effective mass has previously been discredited [23].

Further, Alexandrov and Zhao [2] confuse readers by positing *two* models connecting the supposed OIE in carrier mass to the OIE in $T_C$; one via pair breaking (Eq. (3) of Ref. [2]), and the other based on strong electron-phonon interactions. The authors of the Comment dilute their premise by expecting one to believe in both alternatives simultaneously when neither model fits the data.

If there were an OIE in the carrier mass $m_{ab}^*$ of high-$T_C$ superconductors (owing to an electron-phonon interaction beyond the Migdal approximation) it ought to be observable in not only $\lambda_{ab}$ but also in the normal-state resistivity ($\rho_{ab} \propto m_{ab}^*$). Despite numerous studies of the OIE through resistivity measurements, the absence of any reports of an OIE in normal-state resistivity remains unexplained [1].

## C. Selective treatment of data

The polaron model fostered in the Comment finds that the OIE in the effective mass scales with $\alpha_O$. Hence, the authors of the Comment cite claims that $\delta\lambda_{ab}/\lambda_{ab}$ is proportional to $\alpha_O$. As we showed in Ref. [1], this reading of experiment is premised on selective consideration of only samples with large depressions in $T_C$, large $\Delta T_C$, and Meissner fractions reduced below 20%, i.e., for samples of problematic quality. Experiments actually show that the OIE coefficient for the penetration depth remains the same among samples for $T_C \geq T_{C0}/2$ (fact #3 and Fig. 3; see also Fig. 3 of Ref. [1]). Our pair-breaking model discussed in Ref. [1] provides the natural explanation for the observed invariance in $\delta\lambda_{ab}/\lambda_{ab}$, while polaron theory fails to do so.

## D. Required *d*-wave symmetry

To produce adequate fits of polaron theory to the variation of $\alpha_O$ with $T_C$ observed in the cation-substitution experiments, the method cited in the Comment found that *d*-wave band symmetry must be included [24]. This requirement and discussions of *d*-wave gap symmetry in the Comment conflict with fact #8 in Sect. I. As shown by various bulk probes, the temperature dependence of the penetration depth obeys the two-fluid model, which indicates nodeless (consistent with *s*-wave) pairing symmetry and strong coupling. We refer to NMR work in the pnictides [8], muon spin rotation ($\mu^+$SR) for single crystals (with proper accounting for temperature-activated fluxon pinning) [10] and for high quality polycrystals and heavily twinned crystals (in which fluxons are strongly pinned) [11,12,13], mutual inductance [14], and susceptibility [15]. While our isotope effect analysis itself [1] places no specific requirement on the pairing state symmetry, the observed nodeless behavior (see also Ref. [16] and [17]) is inconsistent with the theoretical arguments and requirements for *d*-wave symmetry put forth by Alexandrov and Zhao in their Comment [2,24].

## IV. Summary and Conclusion

As we discussed in Ref. [1], phonon interactions (as measured by the OIE) increase as $T_C$ decreases below $T_{C0}$ and the Meissner fraction tends to zero. The obvious interpretation of this behavior is that increased phonon interactions indicate a poor quality high-$T_C$ superconductive state and likely help suppress superconductivity in high-$T_C$ materials.

In our work [1] we formulated a pair-breaking model with the added recognition of the limitations of off-stoichiometric sample growth with cation substitutions. In addition to detailing the arguments against phonon-based pairing schemes, we corrected the misrepresentations regarding Pr-doped $YBa_2Cu_3O_{7-\delta}$, showing that the depression of $T_C$ and concomitant increase in the OIE are due to $Pr^{+3}$-on-$Ba^{+2}$-site defects. We showed that the OIE scales with the pair-breaking parameter and depends on whether the cation substitution is *isovalent* or *heterovalent*, with the latter inducing a comparatively greater OIE for the same suppression in $T_C$: Changes in $T_C$ and $\alpha_O$ are much greater for heterovalent substitutions, such as $Pr^{+3}$ for $Ba^{+2}$ as shown in Fig. 2. Our model presented in Ref. [1] treats both valency types consistently in terms of pair breaking. Moreover, the *absence of any Cu d-band signature* in the temperature dependence of the penetration depth, in combination with fact #4, indicates that the superconducting hole condensate is not associated with the $CuO_2$ planes or CuO layers, and hence must reside in the BaO layers; high-$T_C$ superconductivity thus involves at least two bands of carriers, e.g., in the case of $YBa_2Cu_3O_{6.95}$ these carriers are associated with the BaO (holes) and cuprate (electrons) layer structures.

In contrast, the approach taken by Alexandrov and Zhao in their Comment [2] rests upon the faulty assumption that intrinsic superconducting properties such as $T_{C0}$, $\lambda_L$, and $\xi_0$ may be continuously varied by substituting cation impurities into optimum compounds, and yet still accurately reflect the pure superconducting state. It is clear from experiment, however, that such doping compromises the quality of the superconducting state (see fact #2). Moreover, carefully reported experiments caution that large enhancements in $\alpha_O$ correlate with increased $\Delta T_C$, indicative of problems with sample quality [4] (see Figs. 1 and 2). Thus, measurements made on non-stoichiometric materials do not, in general, provide an accurate intrinsic picture of the superconductivity, but are more reflective of the defect structure. Unfortunately, glossing over sample inhomogeneity when interpreting the OIE in $T_C$ and the magnetic penetration depth (or any intrinsic property) is far too common, and conclusions based on data acquired on such samples are inherently suspect [3]. As a consequence, the Fröhlich electron-phonon interaction model of hole pairing advocated by Alexandrov and Zhao [2] (or any phonon-mediated pairing scheme) is rendered invalid, as it contradicts the facts discussed in great detail in our paper [1] and in Sect. I herein. Moreover, the Comment focuses only on the near-isovalent substitution of Zn for Cu, and ignores heterovalent substitutions.

In their critique the authors of the Comment [2] choose to ignore all of the information provided in Ref. [1] and assume incorrectly that these materials can be continuously doped without affecting the quality of the superconducting state. Accepting the reality of inherent problems with the cation-substituted cuprates leads to the pair-breaking formalism of Ref. [1] and exemplified by Eq. (1) (which implicitly satisfies Anderson's theorem since cation doping does not change $T_{C0}$) and the understanding that cation-substituted high-temperature superconductors should not be treated as conventional homogeneously alloyed superconductors. By logical extension of the "virtually unscreened" optical phonons argument advocated by Alexandrov and Zhao [2] to include the doped compounds, one must expect $\alpha_O$ to remain nearly constant (and much larger than observed), in direct contradiction with experiment, nullifying their Comment's "Coulomb's law" critique. Finally, in their penultimate paragraph Alexandrov and Zhao equivocate on the strength of $\lambda$ in a vain attempt to retain relevance of their polaron theory. Thus, the criticisms, premises, arguments and conclusions as put forth by Alexandrov and Zhao in their Comment [2] are without merit.

## Acknowledgments

We are grateful for the support of the U.S. Army Research Office (W911NF-05-1-0346 ARO), Physikon Research Corporation (Project No. PL-206), and the New Jersey Institute of Technology. Publication on this work has appeared [25].